\documentclass{ws-procs9x6-cpt22}
\begin{document}

\newcommand{\refeq}[1]{(\ref{#1})}
\def\etal {{\it et al.}}

\title{Probing Gravity for one Minute\\ with an Optical-Lattice Atom Interferometer}

\author{C.D.\ Panda, M.\ Tao, J.\ Eggelhof, M.\ Ceja, A.\ Reynoso, V.\ Xu, and H. \ M{\"u}ller}

\address{University of California,
Berkeley, CA 94720, USA}



\begin{abstract}
We have realized an atom interferometer that probes gravitational potentials by holding, rather than dropping, atoms. Up to one minute of coherence times are realized by suspending the spatially separated atomic wave packets in an optical lattice that is mode-filtered by an optical cavity. This trapped configuration suppresses phase variance due to vibrations by four to five orders of magnitude, overcoming the dominant noise source in atom-interferometric gravimeters. Recent progress in characterizing and reducing interferometer decoherence led to major increases in coherence and precision, paving the way to measurements of dark-energy candidates and probes of the quantum nature of gravity through measuring the gravity of source masses with record precision and spatial resolution. 
\end{abstract}

\bodymatter
\phantom{}\vskip10pt\noindent
Light-pulse atom interferometry\cite{Cronin2009} has been used to measure the fine structure constant,\cite{Morel2020} the gravitational constant,\cite{Rosi2014} to test the equivalence principle,\cite{Asenbaum2020} to look for laboratory signatures of dark energy,\cite{Jaffe2017} or to search for CPT or Lorentz violation.\cite{Chung2009,Bailey2006} Atom interferometers are among the most accurate and sensitive tools for measuring gravity, in and out of the laboratory.\cite{Tino2021} The achieved precision is limited by two main factors: coherence time and sensitivity to vibrations. In atom interferometers performed with atomic fountains, coherence is limited by the available free-fall time to below three seconds (for $\sim10\,$m drop towers).

\begin{figure}[t]
\begin{center}
\includegraphics[width=\textwidth]{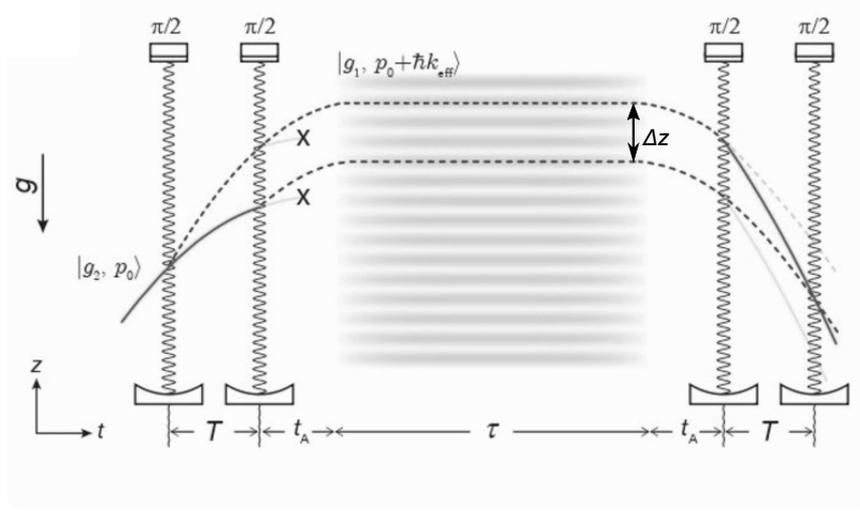}
\end{center}
\caption{\textbf{Lattice interferometer atom trajectories.} Ultracold Cesium atoms are launched against gravity. A pair of  $\pi$/2 pulses creates a spatial superposition state where two atomic wavepackets are separated by a distance $\Delta z$. After a time $t_{\textrm A}$, atoms reach the apex of their free-fall trajectory and are loaded in a far-detuned optical lattice formed by the mode of an optical cavity (horizontal stripes) and remain held for a time $\tau$. The two wavepackets are recombined and their interference is measured.}
\label{fig1}
\end{figure}

\begin{figure}[t]
\begin{center}
\includegraphics[width=\textwidth]{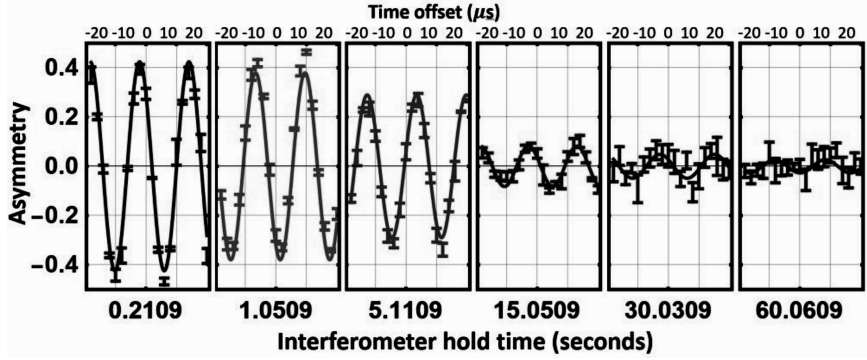}
\end{center}
\caption{One-minute interference fringes measured with the lattice interferometer.}
\label{fig2}
\end{figure}

Our interferometer suspends atoms against Earth's gravity in an optical lattice that is formed by the mode of an optical cavity (Fig.~\ref{fig1}). We observe up to 1 minute of coherence time (Fig.~\ref{fig2}), which is by far the longest-lasting realization of the fundamental notion that a massive quantum particle can be in a superposition of being located in two different places at once and still three times longer than we have realized with the same experiment in a previous publication.\cite{Xu2019} Achieving this in a conventional free-fall atom interferometer would require a drop tower $4.5\,$km tall. 

Atoms held at fixed locations can measure extremely weak fields, such as the gravity due to small milligram source masses, for many seconds. This is orders of magnitude longer than possible with atomic fountains and represents a major sensitivity boost for fundamental-physics experiments, such as tests of fifth forces,\cite{Wolf2007} searches for dark energy in the laboratory,\cite{Jaffe2017} proposed measurements of the gravitational Aharonov--Bohm effect, \cite{Hohensee2012} or tests on whether gravity can mediate entanglement between quantum mechanical systems.\cite{Carney2021} The optical-lattice interferometer is also significantly less susceptible to noise due to acousto-mechanical vibrations than its fountain counterpart. Holding the atoms in an optical lattice for 60 seconds provides $10^4$--$10^5$ times suppression of environmental vibrations in the $1$--$1000\,$Hz vibration band, making operation near the standard quantum limit possible.

Bringing optical-lattice technology, common to atomic-clock metrology,\cite{Marti2018} to atom interferometry is nontrivial, as spatial-superposition states are particularly susceptible to external forces. The interaction of the atoms with the optical lattice causes shifts of the atomic partial wavepacket phases that are large compared to the measured signals. This common-mode phase is intrinsically subtracted by the interferometer, but any imperfections in the optical potential can lead to phase shifts and decoherence. Precise control of the optical field is therefore necessary, which we achieve at least partially by using an optical cavity as a mode-filter for the optical lattice.

\begin{figure}[t]
\begin{center}
\includegraphics[width=\textwidth]{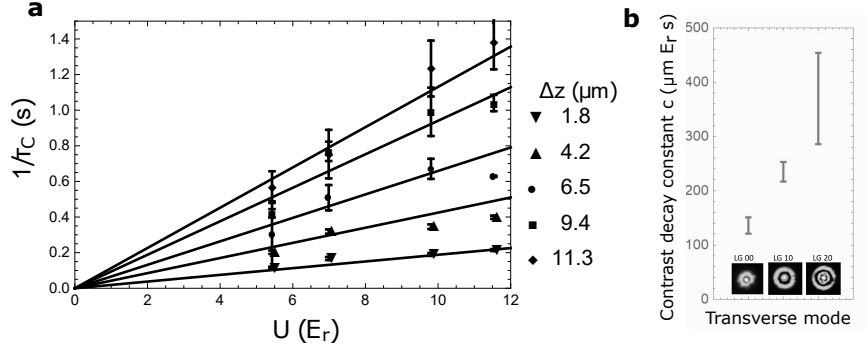}
\end{center}
\caption{\textbf{Interferometer coherence.} \textbf{a.)} Measured linear scaling of $1/\tau_C$ with trap depth $U$ and separation, $\Delta z$.  \textbf{b.)} The magnitude of the contrast decay constant $c$ increases when the optical-lattice radius is lowered through the use of higher-order Laguerre--Gaussian modes of the cavity.
}
\label{fig3}
\end{figure}

We empirically observe a robust scaling in the decoherence rate of our interferometer, where the contrast follows an exponential decay, $C=C_0\exp(-\tau/\tau_C)$ with $C_0$ the starting contrast. $\tau_C$ is inversely proportional to the wavepacket separation $\Delta z$ and trap depth $U$: $1/\tau_C=U \Delta z/c$, where $c=110 \pm30~\mu\textrm{m}\,E_r\,\textrm{s}$ 
is a measured constant and $E_r$ is the recoil energy of the Cs D2 transition (Fig.~\ref{fig3}a). The decoherence results from residual transverse motion of the atoms in the one-dimensional optical lattice, where atoms sample tiny differences in the optical potential holding the top and bottom partial wavepackets. We confirm this by observing increased contrast when only selecting cold atoms by loading atoms in the narrower higher-order Laguerre--Gaussian modes of the cavity (Fig.~\ref{fig3}b). Planned reductions in the atom sample temperature and control of lattice imperfections promise to increase coherence times past the one-minute mark.

We have described an optical-lattice atom interferometer with coherence times reaching one minute, realizing the longest spatial superposition states to date, more than an order of magnitude longer than atomic-fountain interferometers. This geometry comes along with intrinsic advantages: vibration-noise rejection and micron-scale spatial resolution of the measured signals. Cognizance and reduction of decoherence and atom-source upgrades promise to push this technology to record precision and ultra-long coherence. This opens a new era in atom interferometry, with applications in gravimetry, searches for fifth forces, a measurement of the gravitational Aharonov--Bohm effect in the absence of forces, or new tests of CPT and Lorentz violation.


\begin{thebibliography}{xx}

\bibitem{Cronin2009}
A.D.\ Cronin, J.\ Schmiedmayer, and D.E.\ Pritchard,
Rev.\ Mod.\ Phys.\ \textbf{81}, 1051 (2009).



\bibitem{Morel2020}
L.\ Morel, Z.\ Yao, P.\ Clade, and S.\ Guellati-Khélifa, Nature \textbf{588}, 61 (2020).

\bibitem{Rosi2014}
G.\ Rosi, F.\ Sorrentino, L.\ Cacciapuoti, M.\ Prevedelli, and G.M.\ Tino, Nature \textbf{510}, 518 (2014).


\bibitem{Asenbaum2020}
P.\ Asenbaum, C.\ Overstreet, M.\ Kim, J.\ Curti, and M. A.\ Kasevich, Phys.\ Rev.\ Lett.\ \textbf{125}, 191101 (2020).

\bibitem{Jaffe2017}
M.\ Jaffe 
\etal, Nature Phys.\ 13, \textbf{938} (2017).

\bibitem{Chung2009}
K.\ Chung, S.\ Chiow, S.\ Herrmann, S.\ Chu, and H.\ M{"u}ller, Phys.\ Rev.\ D \textbf{80}, 016002 (2009).

\bibitem{Bailey2006}
Q.G.\ Bailey and V.A.\ Kosteleck\'y, Phys.\ Rev.\ D \textbf{74}, 045001 (2006).

\bibitem{Tino2021}
G.M.\ Tino, Quantum\ Sci.\ Technol.\ \textbf{6}, 024014 (2021).


\bibitem{Xu2019}
V.\ Xu, M.\ Jaffe, C.D.\ Panda, S.L.\ Kristensen, L.W.\ Clark, and H.\ M{\"u}ller, Science \textbf{366}, 745 (2019).

\bibitem{Wolf2007}
P.\ Wolf, P.\ Lemonde, A.\ Lambrecht, S.\ Bize, A.\ Landragin, and A.\ Clairon, Phys.\ Rev.\ A \textbf{75}, 063608 (2007).

\bibitem{Hohensee2012}
M.A.\ Hohensee, B.\ Estey, P.\ Hamilton, A.\ Zeilinger, and H.\ M{\"u}ller, Phys.\ Rev.\ Lett.\ \textbf{108}, 230404 (2012).


\bibitem{Carney2021}
D.\ Carney, H.\ M{\"u}ller, and J.M.\ Taylor, PRX Quantum \textbf{2}, 030330 (2021).

\bibitem{Marti2018}
G.E.\ Marti, R.B.\ Hutson, A.\ Goban, S.L.\ Campbell, N.\ Poli, and J.\ Ye, Phys.\ Rev.\ Lett. \textbf{120}, 103201 (2018).


\end{thebibliography}
\end{document}